\documentstyle[aps,prb,epsf]{revtex}
\begin{document}
\newcommand{\bn}{{\bf n}}
\newcommand{\bp}{{\bf p}}
\newcommand{\br}{{\bf r}}
\newcommand{\bq}{{\bf q}}
\newcommand{\bj}{{\bf j}}
\newcommand{\eps}{\varepsilon}
\newcommand{\la}{\langle}
\newcommand{\ra}{\rangle}
\newcommand{\cK}{{\cal K}}
\newcommand{\cD}{{\cal D}}
\large

\title{
  Long-Range Coulomb Interaction and the Crossover 
  between Quantum and Shot Noise in Diffusive Conductors 
}
\author{
         K.\ E.\ Nagaev
}
\address{
  Institute of Radioengineering and Electronics,
  Russian Academy of Science, Mokhovaya 11,
  Moscow 103907, Russia
}
\maketitle
\bigskip
\begin{abstract}
Frequency-dependent nonequilibrium noise in quantum-coherent
diffusive conductors is calculated with account taken of 
long-range Coulomb interaction. For long and narrow contacts
with strong external screening the crossover between quantum and
shot noise takes place at frequencies much smaller than the
voltage drop across the contact. We also show that under certain
frequency limitations, the semiclassical and quantum-coherent
approaches to shot noise are mathematically equivalent.
\end{abstract}

\section{INTRODUCTION}

\par
Recently, the shot noise in mesoscopic contacts became a subject
of extensive study (see recent book \cite{Kog-book} and review
\cite{deJong-97}). 
The 1/3 shot noise suppression in diffusive metal contacts was
obtained almost simultaneously in 1992 using the
quantum-coherent \cite{Beenakker-92} and semiclassical
\cite{Nag-92} approaches. Since then it was disputed whether
both approaches addressed the same effect. For example, de Jong
and Beenakker \cite{deJong-97} suggested that quantum coherence
is not required for the 1/3 suppression, while Landauer
\cite{Landauer} argued that semiclassical and quantum-coherent 
effects are different and 1/3 is just a numerical coincidence.
Although very important properties of shot noise like
universality \cite{Nazarov-94} and exchange effect in
multiterminal contacts \cite{Blanter-97} originally obtained
within the quantum-coherent approach were then rederived using
the semiclassical approach,\cite{Sukhorukov-98} a direct proof
of equivalence of these approaches is still absent. An important
step in this direction was very recently made by Gramespacher
and B\"{u}ttiker, who introduced effective local distribution
function for a phase-coherent nonequilibrium conductor and
expressed the current fluctuations at local tunnel probes
attached to it in terms of this function.
\cite{Gramespacher-99}

\par
A related problem is the frequency dependence of nonequilibrium
noise in diffusive contacts. This dependence is affected by
quantum effects and effects of self-consistent electrical field.
While the former effects were successfully described in terms of
quantum-coherent formula for the shot noise,\cite{Buttiker-92}
the latter presented a certain difficulty for this approach (see
\cite{Buttiker-96} for a more detailed discussion of this
issue). Although particular results were obtained for
nonequilibrium charge fluctuations in quantum point contacts and
chaotic cavities in terms of discrete
capacitances,\cite{Pedersen-98} they could not be extended to
metallic diffusive conductors, for which the potential needs to
be treated as a field. On the contrary, the semiclassical
approach provides an easy treatment of long-range
electron-electron interaction \cite{Naveh-97,Nag-98} yet rules
out the description of quantum effects.

\par
This paper is intended to some extent to bridge the gap between the
phase-coherent and semiclassical approaches. We present a method
for calculating frequency-dependent shot noise in a
phase-coherent diffusive metallic conductor with account taken
of self-consistent electrical field. This method is an extension
of Keldysh Green's function formalism used by Altshuler {\it et
al.} \cite{Altshuler-94} to the case of interacting
electrons.\cite{Kogan-91} We show that phase-coherent equations
may be written in the form very similar to the semiclassical
ones, so that both approaches are {\it mathematically}
equivalent at sufficiently low frequencies. As an illustration
of this method, we calculate quantum noise in nonequilibrium
contacts in the presence of a strong external screening.

\section{BASIC EQUATIONS}

\par
To calculate the noise, we make use of Keldysh diagrammatic
technique, which is valid not only at finite temperatures, but
also for nonequilibrium systems. The spectral density of noise
is calculated as the Fourier transform of symmetrized
current-current correlator
\begin{equation}
  S_{\alpha\beta}(\br_1, \br_2, \omega)
  =
  \int d(t_1 - t_2) \exp[i\omega(t_1 - t_2)]
  \left\la\left\la
    \delta\hat{j}_{\alpha} ( \br_1, t_1 )
    \delta\hat{j}_{\beta}  ( \br_2, t_2 ) 
    +
    \delta\hat{j}_{\beta}  ( \br_2, t_2 )
    \delta\hat{j}_{\alpha} ( \br_1, t_1 )
  \right\ra\right\ra,
\label{S-1}
\end{equation}
where the double angular brackets denote quantum-mechanical and
statistical averaging and 
$\delta\hat{\bf j} = \hat{\bf j} - \la\la\hat{\bf j}\ra\ra$.
In terms of Keldysh contour,\cite{Keldysh-64} this expression
may be written in the form
$$
  S_{\alpha\beta}(\br_1, \br_2, \omega)
  =
  \frac{ e^2 }{ 4m }
  \left(
    \frac{ 
      \partial 
    }{ 
      \partial\br_{1\alpha}
    }
    -
    \frac{
      \partial 
    }{ 
      \partial\br_{1\alpha}'
    }
  \right)
  \left(
    \frac{ 
      \partial 
    }{ 
      \partial\br_{2\beta}
    }
    -
    \frac{ 
      \partial 
    }{ 
      \partial\br_{2\beta}'
    }
  \right)
$$ $$
  \times
  \int d(t_1 - t_2) \exp[i\omega(t_1 - t_2)]
  \bigl\la
      G_{2121}^{II}( 
              \br_1,  t_1 + \delta;
              \br_2,  t_2 - \delta;
              \br_1', t_1 - \delta;
              \br_2', t_2 + \delta
      )
$$ $$
      +
      G_{1212}^{II}(
              \br_1,  t_1 - \delta;
              \br_2,  t_2 + \delta;
              \br_1', t_1 + \delta;
              \br_2', t_2 - \delta
      )
      -
      G_{22}(
              \br_1,  t_1 + \delta;
              \br_1', t_1 - \delta
      )
      G_{11}(
              \br_2,  t_2 - \delta;
              \br_2', t_2 + \delta
      )
$$  \begin{equation}
      -
      G_{11}(
              \br_1,  t_1 - \delta;
              \br_1', t_1 + \delta
      )
      G_{22}(
              \br_2,  t_2 + \delta;
              \br_2', t_2 - \delta
      )  
  \bigr\ra, 
\label{S-2}
\end{equation} 
where $\br_1' \to \br_1$ and $\br_2' \to \br_2$. The two-particle
Green's functions are given by the expression 
$$
  G_{ijkl}^{II}(\br_1 t_1; 
                \br_2 t_2; 
                \br_3 t_3; 
                \br_4 t_4
  )
  =
  \left\la\left\la
     T_c
     \psi    ( \br_1, t_{1i} )
     \psi    ( \br_2, t_{2j} )
     \psi^{+}( \br_3, t_{3k} )
     \psi^{+}( \br_4, t_{4l} )
  \right\ra\right\ra
$$
and $T_c$ is the operator of time ordering on the two-branch
Keldysh temporal contour. Indices $i,j,k,l$ denote the number
of contour branch. It is equal to 1 if the branch goes from
$-\infty$ to $+\infty$ and 2 if it goes from $+\infty$ to
$-\infty$. The points of the first branch are considered as
preceding to the points of the second branch. 
\begin{figure}[t]
\epsfxsize9cm
\centerline{\epsffile{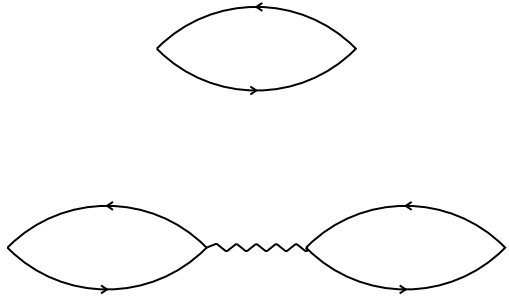}}
\caption{%
 Basic diagrams for the spectral density of noise.
}
\label{fig_1}
\end{figure}

\par
The diagrams
contributing to $ S_{\alpha\beta}$ in the random-phase
approximation are shown in Fig. 1. Each vertex
corresponds to one of the branches of the Keldysh contour,
solid lines correspond to single-electron Green's functions 
$$
  G_{ij}(\br_1 t_1, \br_2 t_2)
  =
  -i
  \left\la\left\la
    T_c
    \psi    ( \br_1 t_{1i} )
    \psi^{+}( \br_2 t_{2j} )
  \right\ra\right\ra,
$$
and the wavy line corresponds to the potential of screened
Coulomb interaction $V_{ij}$. The summation over the Keldysh
indices is performed at the inner vertices. The diagrams shown
in Fig. 1 have a common property that the small momentum
$q_{tr}$ of the order of inverse contact length $L^{-1}$ is
transferred from the left to the right through a single channel,
i.e. by a single interaction line or a pair of lines describing
electron--hole propagation.  The diagrams that correspond to a
redistribution of this momentum among different interaction
lines or interaction lines and electron--hole pairs involve
additional integrations over the momentum transfer and
contain a small factor $1/p_F l_{imp}$, where $p_F$ is the Fermi
momentum and $l_{imp}$ is the elastic mean free path of
electron.\cite{Altshuler-85}

\par
The four functions $G_{ij}$ may be expressed in terms of the
retarded Green's function $G^R$, advanced Green's function
$G^A$, and the electron correlator $G^K$:
$$
  G_{11} = ( G^R + G^A + G^K )/2,
\qquad
  G_{12} = (-G^R + G^A + G^K )/2,
$$ $$
  G_{21} = ( G^R - G^A + G^K )/2,
\qquad
  G_{22} = (-G^R - G^A + G^K )/2.
$$
Similarly, the four potentials $V_{ij}$ may be expressed in
terms of the three independent quantities
$$
  V_{11} = ( V^R + V^A + V^K )/2,
\qquad
  V_{12} = ( V^R - V^A - V^K )/2,
$$ $$
  V_{21} = (-V^R + V^A - V^K )/2,
\qquad
  V_{22} = (-V^R - V^A + V^K )/2.
$$
On this substitution, the spectral density assumes the form
$$
  S_{\alpha\beta}(\br_1, \br_2, \omega)
  =
  S_{\alpha\beta}^{(0)}
  -
  \frac{i}{4}
  \int d^3 r_3 \int d^3 r_4
  \left[
    Q_{\alpha}(\br_1, \br_3, \omega)
    V^R(\br_3, \br_4, \omega)
    P_{\beta}(\br_2, \br_4, -\omega)
  \right.
$$   \begin{equation}
  \left.
    +
    P_{\alpha}(\br_1, \br_3, \omega)
    V^A(\br_3, \br_4, \omega)
    Q_{\beta}(\br_2, \br_4, -\omega)
    +
    Q_{\alpha}(\br_1, \br_3, \omega)
    V^K(\br_3, \br_4, \omega)
    Q_{\beta}(\br_2, \br_4, -\omega)
  \right].
\label{S-3}
\end{equation}
The quantity
$$
  S_{\alpha\beta}^{(0)}( \br_1, \br_2, \omega)
  =
  -
  \frac{1}{16\pi}
  \frac{ e^2 }{ m^2 }
  \left(
    \frac{ 
      \partial 
    }{ 
      \partial\br_{1\alpha}
    }
    -
    \frac{
      \partial 
    }{ 
      \partial\br_{1\alpha}'
    }
  \right)
  \left(
    \frac{ 
      \partial 
    }{ 
      \partial\br_{2\beta}
    }
    -
    \frac{ 
      \partial 
    }{ 
      \partial\br_{2\beta}'
    }
  \right)
  \int d\eps
  \left\la
    G^R( \br_1', \br_2, \eps + \omega )
    G^A( \br_2', \br_1, \eps )
  \right.
$$       \begin{equation}
  \left.
    +
    G^A( \br_1', \br_2, \eps + \omega )
    G^R( \br_2', \br_1, \eps )
    +
    G^K( \br_1', \br_2, \eps + \omega )
    G^K( \br_2', \br_1, \eps )
  \right\ra
\label{S_0-1}
\end{equation}
is the spectral density of noise of noninteracting electrons, 
and
$$
  P_{\alpha}( \br_1, \br_2, \omega )
  =
  -
  \frac{ ie }{ 2m }
  \left(
    \frac{ 
      \partial 
    }{ 
      \partial\br_{1\alpha}
    }
    -
    \frac{
      \partial 
    }{ 
      \partial\br_{1\alpha}'
    }
  \right)
  \int 
  \frac{ d\eps }{ 2\pi }
  \left\la
    G^R( \br_1', \br_2, \eps + \omega )
    G^A( \br_2', \br_1, \eps )
  \right.
$$              \begin{equation}
  \left.
    +
    G^A( \br_1', \br_2, \eps + \omega )
    G^R( \br_2', \br_1, \eps )
    +
    G^K( \br_1', \br_2, \eps + \omega )
    G^K( \br_2', \br_1, \eps )
  \right\ra
\label{P-1}
\end{equation}
is the current - density correlator of noninteracting electrons.
Single angular brackets denote impurity averaging. Both
quantities have a similar structure, i.e. they contain an even
number of correlators $G^K$. In contrast to this, the quantity
$$
  Q_{\alpha}( \br_1, \br_2, \omega )
  =
  -
  \frac{ ie }{ 2m }
  \left(
    \frac{ 
      \partial 
    }{ 
      \partial\br_{1\alpha}
    }
    -
    \frac{
      \partial 
    }{ 
      \partial\br_{1\alpha}'
    }
  \right)
  \int 
  \frac{ d\eps }{ 2\pi }
  \left\la
    G^K( \br_1', \br_2, \eps + \omega )
    G^A( \br_2', \br_1, \eps )
  \right.
$$              \begin{equation}
  \left.
    +
    G^R( \br_1', \br_2, \eps + \omega )
    G^K( \br_2', \br_1, \eps )
  \right\ra
\label{Q-1}
\end{equation}
contains only one $G^K$ and presents the current response to the
external electric potential. Hence the first two terms in
(\ref{S-3}) have a simple physical meaning: they represent
correlations between the current fluctuations caused directly by
random motion of noninteracting electrons and the current
fluctuations induced by density fluctuations of noninteracting
electrons through the mediation of electric field. As will be
shown below, the quantity $V^K$ represents a correlator of
electric potentials, and the last term in (\ref{S-3}) presents a
correlator of currents induced by their fluctuations.

\section{EVALUATION OF DIAGRAMS}

\par
The key point in calculating the correlators of noninteracting
electrons and their responses is the impurity averaging of the
corresponding expressions. We start from the expressions for
impurity-averaged single-electron Green's functions. The
retarded and advanced Green's functions are given by expressions
\begin{equation}
  G^R( \br_1 - \br_2 )
  =
  \int
  \frac{ d^3 p }{ (2\pi)^3 }
  \,
  \frac{
    \exp[ 
          i\bp ( \br_1 - \br_2 )
    ]
  }{
    \eps - \xi - e\phi - i/2\tau
  },
\qquad
  G^A( \br_1 - \br_2 )
  =
  \int
  \frac{ d^3 p }{ (2\pi)^3 }
  \,
  \frac{
    \exp[ 
          i\bp ( \br_1 - \br_2 )
    ]
  }{
    \eps - \xi - e\phi + i/2\tau
  },
\label{G^R}
\end{equation}
where $\phi$ is the local electric potential, which only
slightly changes at their decay length $l_{imp}$.  
The Keldysh Green's function, which contains
information about the electron distribution, is given in
equilibrium by
$$
 G^K(\eps)
 =
 \left[
   G^A(\eps) - G^R(\eps)
 \right]
 \tanh(\eps/2T).
$$
In the general case of a nonequilibrium dirty metal, it is
conveniently expressed by means of the Dyson equation
\begin{equation}
  G^K(\eps, \br_1, \br_2)
  =
  \frac{ 1 }{ 2\pi N_F\tau }
  \int d^3 r'
  G^R( \eps, \br_1 - \br' )
  \bar{G}( \eps, \br' )
  G^A( \eps, \br' - \br_2 ),
\label{G^K-1}
\end{equation}
where $N_F$ is the Fermi density of states and
$$
 \bar{G}( \eps, \br ) 
 \equiv
  G^K(\eps, \br, \br).
$$
By setting $\br_1 = \br_2$ in eqn. (\ref{G^K-1}), one may obtain a
diffusion equation for $\bar{G}$ in the form \cite{Larkin-86}
\begin{equation}
  D_0\nabla^2\bar{G} = 0,
\label{G^K-2}
\end{equation}
where $D_0 = v_F^2\tau/3$ is the diffusion coefficient. Equation
(\ref{G^K-1}) suggests that all the diagrams for the expressions
(\ref{S_0-1}) - (\ref{Q-1}) are constructed of retarded and
advanced Green's functions including Keldysh function vertices
$(2\pi N_F\tau)^{-1}\bar{G}$ and dressed with the random
impurity potential correlators 
$(2\pi N_F\tau)^{-1}\delta(\br_1 - \br_2)$. Summing up an
impurity-potential ladder between oppositely directed retarded
and advanced Green's functions results in a diffuson
\begin{equation}
  {\cal D}( \br_1, \br_2, \omega)
  =
  \left\la
    G^R( \br_1, \br_2, \eps )
    G^A( \br_2, \br_1, \eps + \omega )
  \right\ra,
\label{D-1}
\end{equation}
which obeys the equation \cite{Altshuler-85}
\begin{equation}
  \left[
    i\omega
    +
    D_0
    \frac{
       \partial^2
    }{
       \partial \br_1^2
    }
  \right]
  {\cal D}( \br_1, \br_2, \omega )
  =
  -2\pi N_F
  \delta( \br_1 - \br_2),
\label{D-2}
\end{equation}
and summing this ladder between the functions of the same
direction results in a Cooperon, which obeys a similar equation.
When performing the averaging, we retained only the terms of
lowest order in $1/p_Fl_{imp}$. This excludes
any integrations over the momentum transferred by diffusons or
Cooperons.  In particular, any Cooperon impurity ladders as well
as diffuson impurity ladders between different electron loops
were omitted. This also implies that the localization
corrections are neglected since they contain a small parameter
$1/p_Fl_{imp}$.\cite{Abrikosov} Hence the diagrams to be
selected should consist of ordinary electron loops and diffusons
connected in series and carrying the same momentum transfer
$q_{tr}$.  Our consideration is also restricted to the diffusion
approximation, i.e. frequencies $\omega$ much smaller than the
inverse elastic scattering time $\tau^{-1}$ and characteristic
length scales much larger 

\begin{figure}[t]
\epsfxsize9cm
\centerline{\epsffile{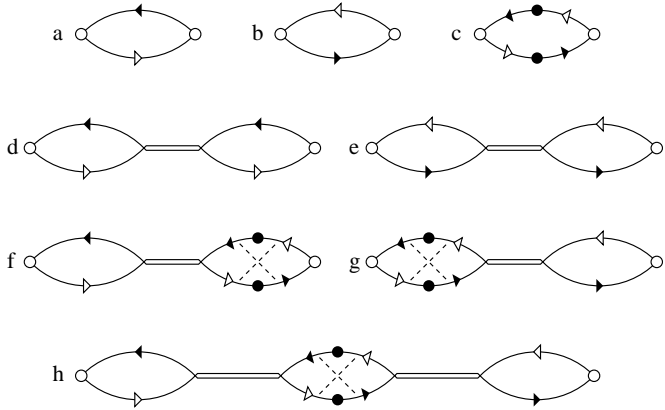}}
\caption{%
 Diagrams for $S_{\alpha\beta}^{(0)}$.
}
\label{fig_2}
\end{figure}
\noindent
than $l_{imp}$. This suggests that in
the momentum representation, the diagrams selected be of lowest
possible order or most singular in $q_{tr}$. 

\par
The diagrammatic expansion for $S_{\alpha\beta}^{(0)}$ is
obtained by collecting diagrams of lowest order in $q_{tr}$
without interaction lines (see Fig. 2). Empty arrows denote
single-electron advanced Green's functions, full arrows denote
retarded Green's functions, full circles denote $(2\pi
N_F\tau)^{-1}\bar{G}$, and empty circles denote current
vertices. The diagrams $a$ - $c$ contain no diffusons. As $G^R$
and $G^A$ decay at a short length $l_{imp}$, they form the term
$\delta$-correlated in space, i.e. the analog of the bare
correlator of extraneous currents in the Langevin approach.

\par
Double lines in diagrams $2d$--$2h$ denote diffusons, which
describe the propagation of electron density fluctuations and
introduce a factor $q_{tr}^{-2}$.  The left current ending to
the diffuson in diagram $2d$ reduces to is derivative $-2\pi
eN_F\tau D_0\partial/\partial r_{1\alpha}$, and the left 
current ending to the diffuson in diagram $2e$, to $2\pi
eN_F\tau D_0\partial/\partial r_{1\alpha}$. The rest of diffuson
endings are given by similar derivatives, their signs depending
on the types of incoming and outgoing Green's functions, and
result in a factor of $q_{tr}$. Physically, these derivatives
reflect the fact that the current fluctuation is proportional to
the density gradient. This is why diagrams $2e$ and
$2d$ appear to be of zeroth order in $q_{tr}$ as well as $2a$ --
$2c$.

\par
The advanced Green's functions change their type for retarded at
the vertices including Keldysh functions $\bar{G}$.  Wherever a
diffuson is interrupted by a pair of such vertices (diagrams
$2f$ -- $2h$), the diffuson frequency changes its sign and a
Hikami box \cite{Hikami-81} denoted by a dashed cross appears.
This box is presented by the sum of the three diagrams shown in
Fig. 3. Though each of them gives a contribution of zeroth order
in $q_{tr}$, they cancel out each other so that the sum vanishes
in a homogeneous metal at $q_{tr} = 0$.  A Hikami box with two
diffuson vertices at $\br_1$ and $\br_2$ is given by the
convolution of
  $(2\pi N_F\tau)^{-1}\bar{G}( \br_3, \eps )$
and
  $(2\pi N_F\tau)^{-1}\bar{G}( \br_4, \eps + \omega)$
with
\begin{equation}
  H_0 (\br_1, \br_2, \br_3, \br_4)
 =
  2\pi N_F \tau^4 D_0
  \delta( \br_1 - \br_2 )
  \delta( \br_1 - \br_3 )
  \delta( \br_1 - \br_4 )
  \sum\limits_{\gamma}
  \left(
    2
    \frac{ \partial }{ \partial r_{1\gamma} }
    \frac{ \partial }{ \partial r_{2\gamma} }
    -
    \frac{ \partial^2 }{ \partial r_{3\gamma}^2 }
    -
    \frac{ \partial^2 }{ \partial r_{4\gamma}^2 }
  \right).
\label{H_0}
\end{equation}
\begin{figure}[b]
\epsfxsize9cm
\centerline{\epsffile{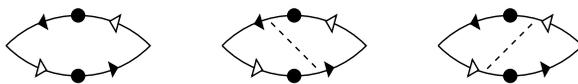}}
\caption{%
 Diagrams for the Hikami box $H_0$. Dashed lines
 represent the correlator of impurity  potential 
 $(2\pi N_F\tau)^{-1}\delta(\br_1 - \br_2)$.
}
\label{fig_3}
\end{figure}
\noindent
and the Hikami box with a current vertex at $\br_1$ and diffuson
vertex at $\br_2$ is given by a convolution of the same
quantities with
\begin{equation}
  H_{\alpha} (\br_1, \br_2, \br_3, \br_4)
  =
  4\pi e N_F \tau^4 D_0
  \delta( \br_1 - \br_2 )
  \delta( \br_1 - \br_3 )
  \delta( \br_1 - \br_4 )
  \frac{ \partial }{ \partial r_{2\alpha} }.
\label{H_a}
\end{equation}
Any diffuson vertex in a Hikami box introduces also a factor of
$(2\pi N_F\tau)^{-1}$.  Correspondingly, these two types of
boxes introduce factors of $q_{tr}^2$ and $q_{tr}$.
Qualitatively, their role can be understood as follows: they
convert local correlators of extraneous Langevin currents
$\delta{j}^{ext}$ at some point into local fluctuations of
charge density $\nabla\delta j^{ext}$, whose propagation to
points $\br_1$ and $\br_2$ is described by the diffusons.

\par
As $\bar{G}$ remain finite at $\eps\to\pm\infty$, integration of
their products with respect to $\eps$ would result in a
divergency. To eliminate these divergencies, diagrams $f$ - $h$
are combined with diagrams $d$ and $e$. Making use of eqn.
(\ref{G^K-2}) and the identity
\begin{equation}
  D_0
  \sum\limits_{\gamma}
  \int d^3 r'
  \frac{
    \partial
    {\cal D} ( \br_1, \br', -\omega )
  }{
    \partial r'_{ \gamma }
  }
  \frac{
    \partial
    {\cal D} ( \br_2, \br', \omega )
  }{
    \partial r'_{ \gamma }
  }
  =
  \pi N_F
  \left[
    {\cal D} ( \br_1, \br_2, -\omega )
    +
    {\cal D} ( \br_2, \br_1, \omega )
  \right],
\label{identity}
\end{equation}
the expression for $S_{\alpha\beta}^{(0)}$ may be presented in
the form
$$
  S_{\alpha\beta}^{(0)}( \br_1, \br_2, \omega)
  =
  4 e^2 N_F D_0
  \delta_{\alpha\beta}
  \delta( \br_1 - \br_2 )
  T_N( \br_1, \omega )
$$ $$
  -
  \frac{2}{\pi} 
  e^2 D_0^2
  \frac{
    \partial^2 {\cal D} ( \br_1, \br_2, -\omega)
  }{
    \partial r_{1\alpha} \partial r_{2\beta}
  }
  T_N( \br_2, \omega )
  -
  \frac{2 }{\pi}
  e^2 D_0^2
  \frac{
    \partial^2 {\cal D} ( \br_2, \br_1, \omega)
  }{
    \partial r_{1\alpha} \partial r_{2\beta}
  }
  T_N( \br_1, \omega )
$$                         \begin{equation}
  +
  \frac{ e^2 D_0^3 }{ \pi^2 N_F }
  \sum\limits_{\gamma}
  \int d^3 r'
  \frac{
    \partial^2 {\cal D} ( \br_1, \br', -\omega)
  }{
    \partial r_{1\alpha} \partial r_{\gamma}'
  }
  T_N( \br', \omega )
  \frac{
    \partial^2 {\cal D} ( \br_2, \br', \omega)
  }{
    \partial r_{\gamma}' \partial r_{2\beta}
  },
\label{S_0-3}
\end{equation}
where
\begin{equation}
  T_N( \br, \omega)
  =
  \frac{ 1 }{ 4 }
  \int d\eps
  \left[
    1
    +
    \frac{ 1 }{ (2\pi N_F)^2 }
    \bar{G}( \br, \eps )
    \bar{G}( \br, \eps + \omega )
  \right]
\label{T_N-1}
\end{equation}
is the effective coordinate- and frequency-dependent noise
temperature.

\par
The most singular diagrams for the correlator $P_{\alpha}$ are
shown in Fig. 4. They are very similar to the diagrams for
$S^{(0)}_{\alpha\beta}$ except that the right-hand current
endings are missing. For this reason, local contributions like
$2a$ -- $2c$ are also missing, because they give a contribution
of the order of $q_{tr}$, 
\begin{figure}[t]
\epsfxsize9cm
\centerline{\epsffile{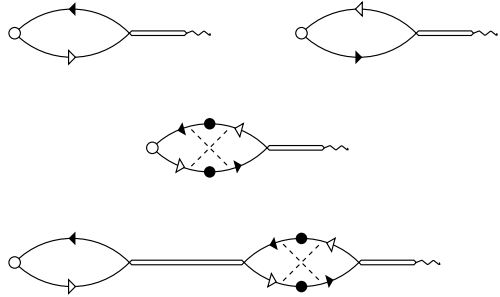}}
\caption{%
 Diagrams for $P_{\alpha}$.
}
\label{fig_4}
\end{figure}
\begin{figure}[t]
\epsfxsize9cm
\centerline{\epsffile{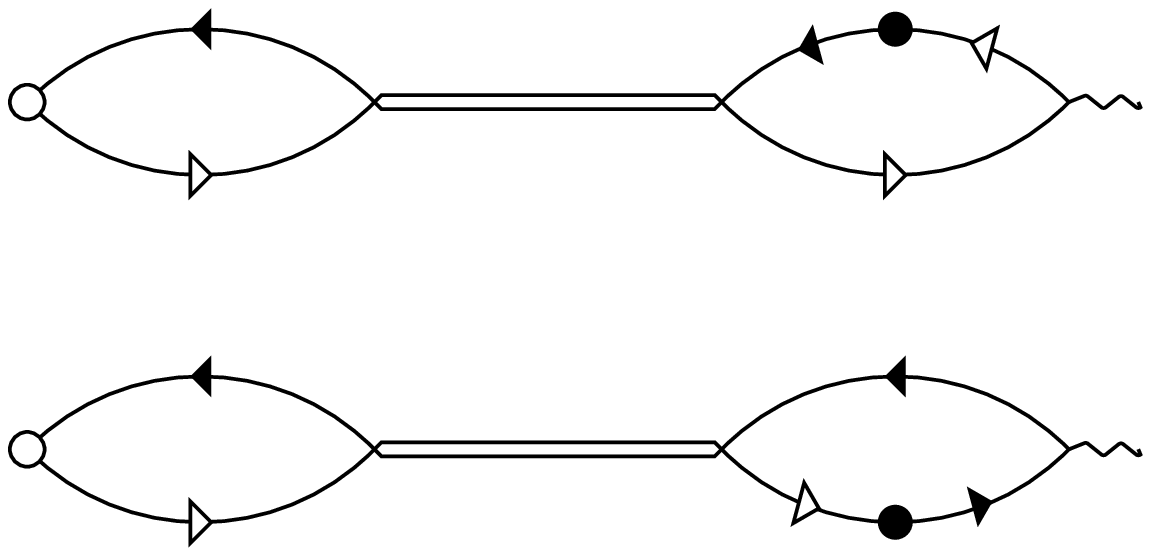}}
\caption{%
 Diagrams for $Q_{\alpha}$.
}
\label{fig_5}
\end{figure}
\noindent
whereas the rest of diagrams are of
the order of $q_{tr}^{-1}$. Combining the two first diagrams
with the three last ones to eliminate the divergencies in
integrals over $\eps$ and making use of the identity
(\ref{identity}), one obtains
$$
  P_{\alpha}( \br_1, \br_2, \omega )
  =
  \frac{4}{\pi}
  eD_0
  \frac{
    \partial {\cal D} ( \br_1, \br_2, \omega)
  }{
    \partial r_{1\alpha} 
  }
  T_N( \br_1, \omega )
$$                       \begin{equation}
  -
  \frac{ 2 e D_0^2 }{ \pi^2 N_F }
  \sum\limits_{\gamma}
  \int d^3 r'
  \frac{
    \partial^2 {\cal D} ( \br_1, \br', -\omega)
  }{
    \partial r_{1\alpha} \partial r_{\gamma}'
  }
  T_N( \br', \omega )
  \frac{
    \partial {\cal D} ( \br_2, \br', \omega)
  }{
    \partial r_{\gamma}' 
  }.
\label{P-2}
\end{equation}
\par
The most singular in $q_{tr}$ diagrams contributing to
$Q_{\alpha}$ are shown in Fig. 5. As this quantity contains only
one Keldysh function, the diagrams for it are essentially
different from those in Figs. 2 and 4. The two terms in (\ref{Q-1})
give contributions of opposite signs yet with the energies in
$\bar{G}$ shifted by $\omega$. Hence the integration over the
energy gives
\begin{equation}
  Q_{\alpha}( \br_1, \br_2, \omega )
  =
  \frac{ \omega }{ \pi }
  eD_0
  \frac{
    \partial {\cal D} ( \br_1, \br_2, -\omega)
  }{
    \partial r_{1\alpha} 
  }.
\label{Q-2}
\end{equation}
Since $Q_{\alpha}$ is proportional to $\omega$, the contribution
from the Coulomb interaction vanishes at $\omega = 0$ and the
noise in this case is essentially the same as for noninteracting
electrons, much like it is in the semiclassical
theory.\cite{Nag-92} 

\par
The potentials of screened interaction satisfy the equations 
\begin{equation}
  V^R ( \br_1, \br_2, \omega )
  =
  V_0 ( \br_1, \br_2 )           
  -
  i\int d^3 \br' \int d^3 \br''
  V_0 ( \br_1, \br' )            
  \Pi^R ( \br', \br'', \omega )
  V^R ( \br'', \br_2, \omega ),
\label{V^R-1}
\end{equation}
\begin{equation}
  V^A ( \br_1, \br_2, \omega )
  =
  V_0 ( \br_1, \br_2 )           
  -
  i\int d^3 \br' \int d^3 \br''
  V_0 ( \br_1, \br' )            
  \Pi^A ( \br', \br'', \omega )
  V^A ( \br'', \br_2, \omega ),
\label{V^A-1}
\end{equation}
and 
$$ 
  V^K ( \br_1, \br_2, \omega )
  =
  -
  i\int d^3 \br' \int d^3 \br''
  V_0 ( \br_1, \br' )            
  \left[
    \Pi^R ( \br', \br'', \omega )
    V^K ( \br'', \br_2, \omega )
  \right.
$$   \begin{equation}
  \left.
    +
    \Pi^K ( \br', \br'', \omega )
    V^A ( \br'', \br_2, \omega )
  \right],
\label{V^K-1}
\end{equation}
where the bare interaction potential $V_0$ satisfies the Poisson
equation
\begin{equation}
  \frac{ 
    \partial^2
  }{
    \partial\br_1^2
  }
  V_0 (\br_1, \br_2 )
  =
  -4\pi e^2
  \delta ( \br_1 - \br_2 ).
\label{V_0}
\end{equation}
The polarization operators 
\begin{equation}
  \Pi^R ( \br_1, \br_2, \omega )
  =
  \frac{ 1 }{ 4\pi }
  \int d\eps
  \left\la
    G^R ( \br_1, \br_2, \eps + \omega )
    G^K ( \br_2, \br_1, \eps )
    +
    G^K ( \br_1, \br_2, \eps + \omega )
    G^A ( \br_2, \br_1, \eps )
  \right\ra,
\label{Pi^R-1}
\end{equation}
\begin{equation}
  \Pi^A ( \br_1, \br_2, \omega )
  =
  \frac{ 1 }{ 4\pi }
  \int d\eps
  \left\la
    G^A ( \br_1, \br_2, \eps + \omega )
    G^K ( \br_2, \br_1, \eps )
    +
    G^K ( \br_1, \br_2, \eps + \omega )
    G^R ( \br_2, \br_1, \eps )
  \right\ra,
\label{Pi^A-1}
\end{equation}
describe the screening of bare potential (\ref{V_0}), and the
polarization operator
$$
  \Pi^K ( \br_1, \br_2, \omega )
  =
  \frac{ 1 }{ 4\pi }
  \int d\eps
  \left\la
    G^R ( \br_1, \br_2, \eps + \omega )
    G^A ( \br_2, \br_1, \eps )
    +
    G^A ( \br_1, \br_2, \eps + \omega )
    G^R ( \br_2, \br_1, \eps )
  \right.
$$  \begin{equation}
  \left.
    +
    G^K ( \br_1, \br_2, \eps + \omega )
    G^K ( \br_2, \br_1, \eps )
  \right\ra.
\label{Pi^K-1}
\end{equation}
actually presents the density -- density correlator of
noninteracting electrons.

\par
Following Altshuler and Aronov,\cite{Altshuler-78} one obtains
expressions for $\Pi^R$ and $\Pi^A$ in the form
\begin{equation}
  \Pi^R ( \br_1, \br_2, \omega )
  =
  \frac{ i }{ 2\pi }
  D_0
  \frac{
    \partial^2 {\cal D} ( \br_1, \br_2, -\omega )
  }{
    \partial \br_1^2
  },
\qquad
  \Pi^A ( \br_1, \br_2, \omega )
  =
  \frac{ i }{ 2\pi }
  D_0
  \frac{
    \partial^2 {\cal D} ( \br_1, \br_2, \omega )
  }{
    \partial \br_1^2
  }.
\label{Pi^R-2}
\end{equation}
For the particular case of an infinite conductor, solving the
integral equations (\ref{V^R-1}) and (\ref{V^A-1}) with the kernels
(\ref{Pi^R-2}) gives the spatial Fourier transforms of $V^R$ and
$V^A$ in the form
\begin{equation}
  V^R ( q, \omega )
  =
  V^{A*} ( q, \omega )
  =
  \frac{
    4\pi e^2
  }{
    q^2 \eps ( q, \omega )
  },
\qquad
  \eps ( q, \omega )
  =
  1
  +
  \frac{
    4\pi\sigma
  }{
    i\omega + D_0q^2
  },
\label{V^R-2}
\end{equation}
where $\sigma = e^2 N_F D_0$ is the conductivity of the metal.

\par
Making use of equations (\ref{V^R-1}) and (\ref{V^A-1}),
the expression for $V^K$ may be recast in the form
\begin{equation}
  V^K ( \br_1, \br_2, \omega )
  =
  -
  i\int d^3 \br' \int d^3 \br''
  V^R ( \br_1, \br', \omega )       
  \Pi^K ( \br', \br'', \omega )
  V^A ( \br'', \br_2, \omega ).
\label{V^K-2}
\end{equation}
This suggests that $V^K$ actually presents a correlator of
fluctuating electric potentials caused by fluctuations of
electron density.

\par
The most singular in $q_{tr}$ diagrams for the density --
density correlator $\Pi^K$ are shown in Fig. 6. Much like the
diagrams for $P_{\alpha}$, these diagrams do not contain local
terms of zeroth order in $q_{tr}$ since the leading contribution
is proportional to $q_{tr}^{-2}$.  The first two diagrams are
combined with the third one using the identity (\ref{identity}),
which results in an analytical expression
\begin{figure}[t]
\epsfxsize9cm
\centerline{\epsffile{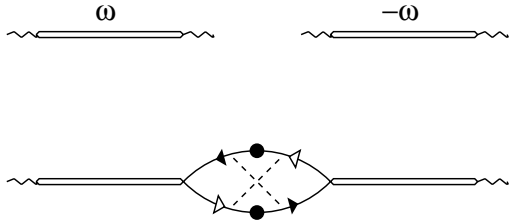}}
\caption{%
 Diagrams for $\Pi^K$.
}
\label{fig_6}
\end{figure}
\begin{equation}
  \Pi^K ( \br_1, \br_2, \omega )
  =
  \frac{ D_0 }{ \pi^2 N_F }
  \int d^3 r'
  \sum\limits_{\gamma}
  \frac{
    \partial {\cal D} ( \br_1, \br', -\omega)
  }{
    \partial r_{\gamma}'
  }
  T_N( \br', \omega )
  \frac{
    \partial {\cal D} ( \br_2, \br', \omega)
  }{
    \partial r_{\gamma}' 
  }.
\label{Pi^K-2}
\end{equation}
It should be noted that the terms involving the electron--electron
interaction in (\ref{S-3}) appear to be more singular in
$q_{tr}$ than $S^{(0)}_{\alpha\beta}$. However they also contain
excess powers of frequency $\omega$, which makes all the terms
in (\ref{S-3}) comparable in the diffusive limit.

\par
In an equilibrium conductor with electric potential $\phi$ the
Green's function $\bar{G}$ is of the form \cite{Larkin-86}
$$
  \bar{G} (\eps)
  =
  -2\pi i N_F
  [
    1
    -
    f_0 ( \eps + e\phi )
  ],
$$
where $f_0 (\eps)$ is the Fermi distribution function for a
given temperature. Hence one may define the local distribution
function in nonequilibrium metal via the relationship
\begin{equation}
  f (\br, \eps )
  =
  \frac{ 1 }{ 2 }
  \left[
    1
    -
    \frac{ i }{ 2\pi N_F }
    \bar{G} ( \br, \eps )
  \right], 
\label{f-1}
\end{equation}
so that it satisfies the standard diffusion equation and the 
boundary conditions for the semiclassical distribution function.
This definition of $f$ is merely equivalent to the definition of
Gramespacher and B\"{u}ttiker in terms of distribution functions
in electrodes and the injectivities of these
electrodes.\cite{Gramespacher-99} In terms of $f$, the effective
noise temperature $T_N$ may be written as
\begin{equation}
  T_N (\br, \omega)
  =
  \frac{ 1 }{ 2 }
  \int d\eps
  \left\{
    f( \br, \eps )
    [
      1 - f ( \br, \eps + \omega )
    ]
    +
    f ( \br, \eps + \omega )
    [
      1 - f ( \br, \eps )
    ]
  \right\}.
\label{T_N-2}
\end{equation}
In equilibrium, this expression reduces to
$$
  T_N
  =
  \frac{1}{2}
  \omega
  \coth
  \left(
    \frac{\omega}{ 2T }
  \right),
$$
which presents the standard factor in the Nyquist formula.

\section{QUANTUM BOLTZMANN -- LANGEVIN SCHEME}

\par
It is easily verified (see Appendix for details) that
calculating the noise on the basis of Eqn. (\ref{S-3}) results
into the expressions for its spectral density identical to those
given by the following quantum extension of the semiclassical
Boltzmann -- Langevin scheme.  As well as in the semiclassical
case, the fluctuation of current is formally described by a
Langevin equation \cite{Nag-98}
\begin{equation}
  \delta{\bf j} 
  =
  -D_0\frac{\partial}{\partial{\bf r}}\,\delta\rho 
  +
  \sigma\,\delta{\bf E} 
  +
  \delta{\bf j^{ext}},
  \label{PRB-5}
\end{equation}
where the fluctuation of charge density $\delta\rho$ satisfies
the equation
\begin{equation}
  \left(
       \frac{\partial}{\partial t} 
       -
       D_0\nabla^2 
       +
       4\pi\sigma
  \right)
  \,
  \delta\rho
  =
  -\nabla\,\delta{\bf j^{ext}},
\label{PRB-7}
\end{equation}
the self-consistent fluctuation of electric potential obeys the
Poisson equation
\begin{equation}
 \delta{\bf E}
 =
 -\nabla\delta\phi,
\qquad
 \nabla^2\delta\phi
 =
 -4\pi\delta\rho,
\label{PRB-10}
\end{equation}
and the spectral density of extraneous currents $\delta j^{ext}$
is given by
 \begin{equation}
 \left\la\left\la
 \delta j_{\alpha}^{ext}({\bf r_1})
 \delta j_{\beta}^{ext}({\bf r_2})
 \right\ra\right\ra_{\omega}
 =
 4\sigma\delta_{\alpha \beta} 
 \delta({\br_1 - \br_2})
 T_N (\br_1, \omega),
 \label{PRB-8}
 \end{equation}
where $T_N$ is expressed in terms of the effective distribution
function (\ref{f-1}). It is noteworthy that the correlator of
these currents is quantum, whereas the response to the
extraneous currents remains semiclassical. The reason for this
is that in dirty metals, the response functions are presented by
diffusons and screened interaction lines connected in series,
which describe classical propagation of fluctuations, and the
interference corrections to them are small. It should be
stressed that these interference corrections have nothing to do
with the quantum noise, which results solely from the shifts in
energy of the Keldysh functions that appear in pairs in ordinary
electron loops and the Hikami boxes. For equilibrium systems,
this noise is a universal feature \cite{Landau} and does not
depend on the presence of interference phenomena.
 
\par
At $\omega \to 0$, equation (\ref{PRB-8})
transforms into semiclassical Eqn. (8) of Ref.\cite{Nag-98}.
Hence it is not surprising that the low-frequency properties of
quantum-coherent shot noise \cite{Nazarov-94,Blanter-97} may be
rederived also in the semiclassical approach.\cite{Sukhorukov-98}

\section{QUANTUM NOISE IN A LONG CONTACT WITH SCREENING}

\par
Now we apply this formalism to calculations of the
zero-temperature shot noise in a long contact with strong
external screening. Consider a contact of length $L$ and
circular section with a diameter $2r_0 \ll L$ that connects two
massive electrodes with a voltage drop $V$ between them. The
third electrode presents a perfectly conducting grounded
shielding of the contact, which is coaxial with it and isolated
from its surface with a thin insulating film of thickness
$\delta_0$ and dielectric constant $\eps_d$ (see Fig. 7, inset).
In this case, the distribution function (\ref{f-1}) obeys a
one-dimensional diffusion equation with the boundary conditions
\begin{equation}
  f(L/2, \eps)
  =
  f_0(\eps - eV/2),
\qquad
  f(L/2, \eps)
  =
  f_0(\eps + eV/2).
\label{f-2}
\end{equation}
At zero temperature, the solution is of the form
\begin{equation}
  f(x, \epsilon) 
  =
  \left\{
  \begin{array}{ll}
    0,       & \epsilon > eV/2\\
    1 - x/L, & eV/2 > \epsilon > -eV/2\\
    1,       & \epsilon < -eV/2.
  \end{array}
  \right.
\label{f-3}
\end{equation}
Hence
\begin{equation}
  T_N (x, \omega)
  =
  \omega/2 
  + 
  \theta( eV - \omega)
  (eV - \omega)
  \frac{ x }{ L }
  \left( 
    1 
    - 
    \frac{ x }{ L }
  \right).
\label{T_N-3}
\end{equation}
Equations (\ref{PRB-5}) -- (\ref{PRB-10}) were solved in
\cite{Nag-98} for the contact geometry in hand.
Using the response of current at the left edge of the contact
$\delta I$ to the extraneous current $\delta j$ obtained there, 
the spectral density of current fluctuations at
one of the contact ends may be presented in the form
\begin{equation}
  S_I( \omega )
  =
  \frac{ 4 }{ RL }
  \int\limits_0^L dx\,
  K( x, \omega ) T_N( x, \omega ),
  \label{S_I-1}
\end{equation}  
where  $R$ is the resistance of the contact, $x$ is the longitudinal
coordinate, and
\begin{equation}
  K( x, \omega )
  =
  2 ( \gamma_{\omega} L )^2
  \frac{
        \cosh [ 2\gamma_{\omega} ( L - x ) ]
        +
        \cos [ 2\gamma_{\omega} ( L - x ) ]
       }{
        \cosh ( 2\gamma_{\omega} L  )
        -
         \cos ( 2\gamma_{\omega} L  )
       },
  \label{K}
\end{equation}
where $\gamma_{\omega} = (\omega\varepsilon_d /
4\pi\sigma\delta_0 r_0)^{1/2}$.  At sufficiently high
frequencies, the kernel $K$ exponentially decreases with $x$.
This decrease has a simple physical explanation. At contact
dimensions much larger than the 
\begin{figure}[t]
\epsfxsize9cm
\centerline{\epsffile{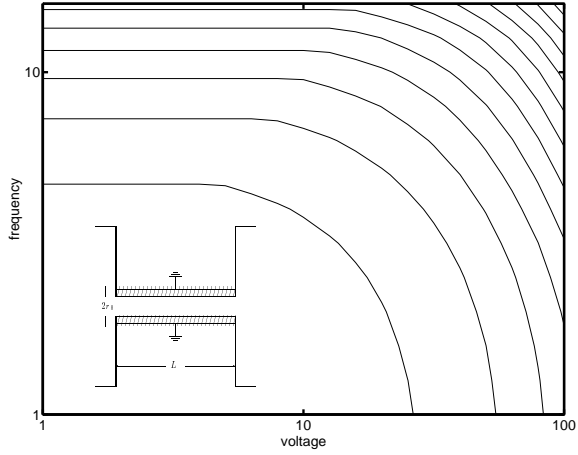}}
\caption{%
  Contour plots of $S_I$ vs. normalized voltage
  $eVL^2/4\pi\sigma\delta_0r_0$ and normalized frequency
$\omega L^2/4\pi\sigma\delta_0r_0$. Inset shows the axial cross
section of the contact.
}
\label{fig_7}
\end{figure}
\noindent
Debye screening length, the
local current fluctuations inside the contact described by
correlator (\ref{PRB-8}) induce the current fluctuations at the
contact edges through the long-range fluctuations of electrical
field.  However at finite frequencies, the field-induced
fluctuational charges are allowed to pile up at the outer
surface of the contact. Hence the electric lines of force
emerging from the middle points of the contact are intercepted
by the screening electrode and do not reach the contact edges.
Hence it is only the portions of the contact adjacent to its
edges that contribute to the measurable
noise.\cite{Naveh-97,Nag-98} The integration gives
$$
  S_I (\omega)
  =
  2
  \frac{ \omega }{ R }
  \gamma_{\omega} L
  \frac{
    \sinh( 2\gamma_{\omega} L )
    +
    \sin( 2\gamma_{\omega} L )
  }{
    \cosh( 2\gamma_{\omega} L )
    -
    \cos( 2\gamma_{\omega} L )
  }        
$$         \begin{equation}         
  +
  2
  \theta ( eV - \omega )
  \frac{ eV - \omega }{ R }
  \left[
    1
    -
    \frac{ 1 }{ \gamma_{\omega} }
  \frac{
    \sinh( 2\gamma_{\omega} L )
    -
    \sin( 2\gamma_{\omega} L )
  }{
    \cosh( 2\gamma_{\omega} L )
    -
    \cos( 2\gamma_{\omega} L )
  }
  \right].
\label{S_I-2}
\end{equation}
In particular, 
\begin{equation}
  S_I
  =
  2
  \frac{ \omega }{ R }
  +
  \frac{ 2 }{ 3 }
  \theta( eV - \omega )
  \frac{ eV - \omega }{ R }
\label{S_I-3}
\end{equation}
at $\gamma_{\omega}L \ll 1$ and
\begin{equation}
  S_I
  =
  2\gamma_{\omega}L
  \frac{ \omega }{ R }
  +
  2
  \theta( eV - \omega )
  \frac{ eV - \omega }{ R }
\label{S_I-4}
\end{equation}
at $\gamma_{\omega}L \gg 1$. Note that in the latter case, the
equilibrium quantum noise grows as $\omega^{3/2}$. In terms of
the Nyquist formula $S_I = 2\omega\coth(\omega/2T){\cal R}{e}\,
Z^{-1}(\omega)$ this superlinear increase is due to the fact
that the real part of inverse effective impedance
$Z^{-1}(\omega)$ of the contact exhibits a square-root increase
in this frequency range.  Contour plots of $S_I$ vs. normalized
voltage and normalized frequency are shown in Fig. 7. From this
figure and Eqn. (\ref{S_I-4}) it is readily seen that quantum
noise (voltage-independent term) dominates over the shot noise
(term proportional to voltage) even at $\omega \ll eV$ provided
that $\omega$ is sufficiently large. The reason is that in this
case the noise is caused by extraneous currents near the contact
edges, where the distribution function only slightly differs
from the equilibrium one.\cite{Nag-92} In the opposite case of a
short contact the standard expression (\ref{S_I-3}) obtained in
Ref.\cite{Buttiker-92} is valid for all frequencies.  Recently,
measurements of quantum shot noise were performed for mesoscopic
diffusive contacts.\cite{Schoelkopf-97,Schoelkopf-98} Hence
these results are of direct experimental interest.

\section{CONCLUSION}

\par
In summary, we have shown that in a phase-coherent diffusive
metal, the high-frequency nonequilibrium noise may be described 
by semiclassical hydrodynamic Langevin equations with quantum
sources. In long contacts with a strong external screening the
crossover between the shot and quantum noise may take place at
frequencies much lower than the voltage drop across the contact
because the noise is dominated by the extreme contact regions,
where the electron distribution is slightly affected by the
applied voltage.

\section{ACKNOWLEDGMENTS}

\par
The author is grateful to G.\ B.\ Lesovik for a stimulating
discussion.
\bigskip
\appendix
\centerline{APPENDIX}

\par\bigskip
The equivalence between the results of diagrammatic approach and
of the proposed quantum extension of Boltzmann -- Langevin
method may be demonstrated as follows.
By applying the Laplace operator to both sides of equations
(\ref{V^R-1}) and (\ref{V^A-1}) and making use of the equation
for the diffuson (\ref{D-2}), one obtains that
$$
  \nabla^2 
  V^R(\br_1, \br_2, \omega)
  =
  -4\pi e^2 \delta(\br_1 - \br_2)
  +
  4\pi e^2N_F 
  V^R(\br_1, \br_2, \omega)
$$ \begin{equation}
  -
  2i\omega e^2
  \int d^3 \br'
  \cD(\br_1, \br', -\omega)
  V^R(\br', \br_2, \omega),
\label{V^R-3}
\end{equation}
$$
  \nabla^2 
  V^A(\br_1, \br_2, \omega)
  =
  -4\pi e^2 \delta(\br_1 - \br_2)
  +
  4\pi e^2N_F 
  V^A(\br_1, \br_2, \omega)
$$ \begin{equation}
  +
  2i\omega e^2
  \int d^3 \br'
  \cD(\br_1, \br', \omega)
  V^A(\br', \br_2, \omega).
\label{V^A-3}
\end{equation}
Using these expressions and the equation for diffuson
(\ref{D-2}), it is easily verified by direct substitution that
the Fourier transform of equation (\ref{PRB-7}) for the
fluctuation $\delta\rho$ is satisfied by 
\begin{equation}
 \delta\rho(\br, \omega)
 =
 \frac{1}{8\pi^2 e^2 N_F}
 \frac{
    \partial^2
 }{
    \partial\br^2
 }
 \int d^3\br' \int d^3\br''
 V^A(\br, \br', \omega)
 \cD(\br', \br'', \omega)
 \nabla\delta\bj^{ext}(\br'').
\label{d-rho}
\end{equation}
In terms of the potential fluctuations $\delta\phi$, the
fluctuation of the current (\ref{PRB-5}) may be presented in the
form
\begin{equation}
 \delta\bj
 =
 -\nabla
 (
   D_0\delta\rho
   +
   \sigma\delta\phi
 )
 +
 \delta\bj^{ext}.
\label{j-1}
\end{equation}
Since the screened potentials $V^{R(A)}$ and the potential
fluctuation $\delta\phi$ obey the same boundary conditions, the
Poisson equation for $\delta\phi$ (\ref{PRB-10}) is easily
solved to give
\begin{equation}
 \delta\phi(\br, \omega)
 =
 -\frac{1}{2\pi e^2 N_F}
 \int d^3\br' \int d^3\br''
 V^A(\br, \br', \omega)
 \cD(\br', \br'', \omega)
 \nabla\delta\bj^{ext}(\br'').
\label{d-phi}
\end{equation}
Making use of identity (\ref{V^A-3}), the Laplace operator may
be excluded from (\ref{d-rho}) and the expression (\ref{j-1})
for $\delta\bj$ may be recast in the form
\begin{equation}
 \delta j_{\alpha}(\br, \omega)
 =
 \delta j_{\alpha}^{ext}(\br, \omega)
 +
 \int d^3\br'
 \cK_{\alpha}(\br, \br', \omega)
 \nabla\bj^{ext}(\br'),
\label{j-2}
\end{equation}
where
\begin{equation}
 \cK_{\alpha}(\br_1, \br_2, \omega)
 =
 \frac{D_0}{2\pi N_F}
 \frac{
    \partial \cD(\br_1, \br_2, \omega)
 }{
   \partial r_{1\alpha}
 }
 -
 \frac{
   i\omega D_0
 }{
   4\pi^2 N_F
 }
 \int d^3 \br' \int d^3 \br''
 \frac{
   \partial \cD(\br_1, \br', \omega)
 }{
   \partial r_{1\alpha}
 }
 V^A(\br', \br'', \omega)
 \cD(\br'', \br_2, \omega).
\label{K_a}
\end{equation}
Integrating in (\ref{j-2}) by parts, multiplying it by the
corresponding equation for $\delta j_{\beta}^{ext}(\br_2,
-\omega)$, and making use of the correlator (\ref{PRB-8}), one
obtains the spectral density of noise in the form
$$
 S_{\alpha\beta}(\br_1, \br_2, \omega)
 =
 4\sigma\delta_{\alpha\beta}\delta(\br_1 - \br_2)
 T_N(\br_1, \omega)
 -
 4\sigma
 \frac{
   \partial\cK_{\beta}(\br_2, \br_1, -\omega)
 }{
   \partial r_{1\alpha}
 }
 T_N(\br_1, \omega)
$$ \begin{equation}
 -
 4\sigma
 \frac{
   \partial\cK_{\alpha}(\br_1, \br_2, \omega)
 }{
   \partial r_{2\beta}
 }
 T_N(\br_2, \omega)
 +
 4\sigma\sum\limits_{\gamma}\int d^3 \br'
 \frac{
   \partial\cK_{\alpha}(\br_1, \br', \omega)
 }{
   \partial r_{\gamma}'
 }
 \frac{
   \partial\cK_{\beta}(\br_2, \br', -\omega)
 }{
   \partial r_{\gamma}'
 }
 T_N(\br', \omega).
\label{S-4}
\end{equation}
Recalling that $V^A(-\omega) = V^R(\omega)$, it is easily seen
that expression (\ref{S-4}) is identical to (\ref{S-3}) with
$S_{\alpha\beta}^{(0)}$, $Q_{\alpha}$, $P_{\alpha}$, and $V^K$
given by (\ref{S_0-3}), (\ref{Q-2}), (\ref{P-2}), and
(\ref{V^K-2}), respectively.



\begin{thebibliography}{99}
%
%
\bibitem{Kog-book} Sh. Kogan, {\it Electronic Noise and 
Fluctuations in Solids}, Cambridge University Press, 1996.
%
\bibitem{deJong-97} 
M. J. M. de Jong and C. W. J. Beenakker, in {\it Mesoscopic
Electron Transport,} NATO ASI, Ser. E, Vol. 345, edited by L. P.
Kouwenhoven, G. Sch\"{o}n, and L. L. Sohn (Kluwer, Dordrecht,
1997). 
%
\bibitem{Beenakker-92}
C. W. J. Beenakker and M. B\"{u}ttiker, Phys. Rev. 
B {\bf 46}, 1889 (1992).
%
\bibitem{Nag-92} 
K. E. Nagaev, Phys. Lett. A {\bf 169}, 103 (1992).
%
\bibitem{Landauer} 
R. Landauer,
Phys.\ Rev.\ B {\bf 47}, 16427 (1993);
Ann.\ N. Y. Acad.\ Sc.\ {\bf 755}, 417 (1995);
Physica B {\bf 227}, 156 (1996).
%
\bibitem{Nazarov-94} 
Yu. V. Nazarov, Phys. Rev. Lett. {\bf 73}, 134 (1994).
%
\bibitem{Blanter-97} 
Ya. M. Blanter and M. B\"{u}ttiker, Phys. Rev. B {\bf 56}, 2127 (1997).
%
\bibitem{Sukhorukov-98}
E. V. Sukhorukov and D. Loss, Phys. Rev. Lett. {\bf 80}, 4959
(1998); Phys. Rev. B {\bf 59}, 13054 (1999).
%
\bibitem{Gramespacher-99}
T. Gramespacher and M. B\"{u}ttiker, Phys. Rev. B {\bf 60}, 2375
(1999).
%
\bibitem{Buttiker-92}  
M. B\"{u}ttiker, Phys. Rev. B {\bf 45}, 3807 (1992).
%
\bibitem{Buttiker-96}  M. B\"{u}ttiker, J. Math. Phys. {\bf 37},
4793 (1996).
%
\bibitem{Pedersen-98} M. H.  Pedersen, S. A. van Langen, and M.
B\"{u}ttiker, Phys. Rev. B {\bf 57}, 1838 (1998).
%
\bibitem{Naveh-97} Y. Naveh, D. V. Averin, and K. K. Likharev, Phys. Rev.
Lett. {\bf 79}, 3482 (1997).
%
\bibitem{Nag-98} K. E. Nagaev, Phys. Rev. B {\bf 57}, 4628 (1998).
%
\bibitem{Altshuler-94} B. L. Altshuler, L. S. Levitov, and A. Yu.
Yakovets, Pis'ma Zh. Eksp. Teor. Fiz.,
{\bf 59}, 821 (1994) [JETP Lett. {\bf 59}, 857 (1994)].
%
\bibitem{Kogan-91} Equations for correlation functions of
Green's function fluctuations were recently derived by Sh. M.
Kogan, Phys. Rev. A {\bf 44}, 8072 (1991). In principle, these
equations also could be used for our purpose. However their
derivation involved a four-branch extension of Keldysh
technique, and we prefer to use its standard version.
%
\bibitem{Keldysh-64} L. V. Keldysh, Zh. Eksp. Teor. Fiz 
{\bf 47}, 1515 (1964) [JETP {\bf 20}, 1018 (1964)].
%
\bibitem{Altshuler-85} B. L. Altshuler and A. G. Aronov, in {\it
Electron-electron Interactions in Disordered Systems,} edited by A.
L. Efros and M. Pollak (North-Holland, Amsterdam, 1985), p. 1.
%
\bibitem{Abrikosov} A. A. Abrikosov, {\it Fundamentals of the
Theory of Metals}, North Holland, Amsterdam, 1988.
%
\bibitem{Larkin-86} A. I. Larkin and D. E. Khmelnitskii, Zh.
Eksp. Teor. Fiz. {\bf 91}, 1815 (1986) [JETP {\bf 64}, 1075
(1986)]. 
%
\bibitem{Hikami-81} S. Hikami, Phys. Rev. B {\bf 24}, 2671
(1981). 
%
\bibitem{Altshuler-78} B. L. Altshuler and A. G. Aronov, Zh.
Eksp. Teor. Fiz., {\bf 75}, 1610 (1978) [Sov. Phys. JETP {\bf
48}, 812 (1978)].
%
\bibitem{Landau} L. D. Landau and E. M. Lifshitz, {\it 
Statistical Physics} (Pergamon, London, 1958).
%
\bibitem{Schoelkopf-97} R. J. Schoelkopf, P. J. Burke, A. A. 
Kozhevnikov, D. E. Prober, and M. J. Rooks, Phys. Rev. Lett.
{\bf 78}, 3370 (1997).
%
\bibitem{Schoelkopf-98} R. J. Schoelkopf, A. A. 
Kozhevnikov, D. E. Prober, and M. J. Rooks, Phys. Rev. Lett.
{\bf 80}, 2437 (1998).
%
\end{thebibliography}
\end{document}